\documentclass[aps,prl,twocolumn,showpacs,superscriptaddress,groupedaddress]{revtex4-1}
\usepackage{graphicx}
\usepackage{amssymb}
\usepackage{amsmath}
\usepackage{latexsym}
\usepackage{ulem}
\usepackage{color}
\usepackage{dcolumn}
\usepackage{subfigure}
\usepackage[T1]{fontenc}
\usepackage[utf8]{inputenc}
\usepackage[english,frenchb]{babel}
\usepackage[colorlinks=true,linkcolor=blue,citecolor=blue]{hyperref}%

\usepackage{bm}        
\usepackage{amssymb}   
\begin{document}

\title{Carrier induced ferromagnetism in the insulating Mn doped III-V semiconductor InP}
\author{Richard Bouzerar}
\email{richard.bouzerar@univ-lyon1.fr}
\affiliation{Univ Lyon, Université Claude Bernard Lyon 1, CNRS, Institut Lumière Matière, F-69622, LYON, France}
\author{Daniel May}
\affiliation{Lehrstuhl f\"ur Theoretische Physic II, Technische Universit\"at Dortmund, 44221 Dortmund, Germany}
\author{Ute  L\"ow}
\affiliation{Lehrstuhl f\"ur Theoretische Physic II, Technische Universit\"at Dortmund, 44221 Dortmund, Germany}
\author{Denis  Machon}
\affiliation{Univ Lyon, Université Claude Bernard Lyon 1, CNRS, Institut Lumière Matière, F-69622, LYON, France}
\author{Patrice  Melinon}
\affiliation{Univ Lyon, Université Claude Bernard Lyon 1, CNRS, Institut Lumière Matière, F-69622, LYON, France}
\author{Shengqiang Zhou}
\affiliation{Helmholtz-Zentrum Dresden-Rossendorf, Institute of Ion Beam Physics and Materials Research, Bautzner Landstrasse 400 01328 Dresden, Germany}
\author{Georges Bouzerar}
\affiliation{Univ Lyon, Université Claude Bernard Lyon 1, CNRS, Institut Lumière Matière, F-69622, LYON, France}

\date{\today}
\selectlanguage{english}
\begin{abstract}
Although InP and GaAs have very similar band-structure their magnetic properties appear to drastically differ. Critical temperatures in (In,Mn)P are much smaller than that of (Ga,Mn)As and scale linearly with Mn concentration. This is in contrast to the square root behaviour found in (Ga,Mn)As. Moreover the magnetization curve exhibits an unconventional shape in (In,Mn)P contrasting with the conventional one of well annealed (Ga,Mn)As. By combining several theoretical approaches, the nature of ferromagnetism in Mn doped InP is investigated. It appears that the magnetic properties are essentially controlled by the position of the Mn acceptor level. Our calculations are in excellent agreement with recent measurements for both critical temperatures and magnetizations. The results are only consistent with a Fermi level lying in an impurity band, ruling out the possibility to understand the physical properties of Mn doped InP within the valence band scenario. The quantitative success found here reveals a predictive tool of choice that should open interesting pathways to address magnetic properties in other compounds.

\end{abstract}
\maketitle

Since their discovery almost two decades ago, III-V based diluted magnetic semiconductors (DMS) rapidly became an intense field of research\cite{satormp,junqwirth,timm,ohno,philip,macdo}.
The hope to incorporate these doped materials in spintronic devices has triggered a competitive race in the search for suitable candidates with sufficiently high Curie temperature.
Since the highest critical temperature has been measured in Mn doped GaAs, this compound became the prototypical III-V DMS. 
Despite all the efforts,  the understanding of carrier induced ferromagnetism in these dilute magnets remains controversial and highly debated. 
In the attempt to explain the observed features of (Ga,Mn)As two antagonist scenari have emerged.
The first is based on the mean field Zener model in which the $pd$-exchange interaction between valence holes and localized Mn-$3d$ electrons is treated perturbatively (valence band (VB) picture). In the second scenario, the Fermi level lies in a completely detached impurity band (impurity band (IB) picture) (see Ref.\cite{junqwirth} for details).
Because most of the available experimental studies mainly focus on Mn doped GaAs, it becomes essential to consider other compounds in order to enlighten the complex underlying physics in these materials.
In recent experimental studies\cite{Zhou-R2014,Khalid-2015} intrinsic ferromagnetism has been reported also in Mn doped InP. In that compound the $T$-dependent magnetization did not show the expected standard Brillouin like shape but rather a rapid linear decrease almost up to $T_C$. Furthermore, the measured critical temperatures were about 3 times smaller than in Mn doped GaAs. Those experimental differences are somehow surprising given that InP and GaAs have very similar band structures, direct band gaps and also the effective masses are very close in both materials. By contrast the most noticeable difference between those two III-V materials is the relative position of the Mn acceptor level, which is $110\,\mathrm{meV}$ in GaAs\cite{Yu} and $220\,\mathrm{meV}$ in InP\cite{Clerjaud}. 
The availability of these recent data, on a compound which differs mainly from (Ga,Mn)As by the position of the impurity acceptor level, allows to isolate the influence of that specific parameter. This is a rare opportunity to test the broad scope of the model Hamiltonian presented below. In this letter, we are shedding some light on the nature of the carrier induced ferromagnetism in Mn doped InP and provide both qualitative and quantitative theoretical understanding of the features observed in recent experimental studies.

To address this issue theoretically, we proceed within a two step procedure described as follows. Starting from a minimal model Hamiltonian, the first step is devoted to the calculation of the carrier mediated Mn-Mn magnetic couplings. In the second step we calculate the magnetic properties of the derived dilute effective Heisenberg Hamiltonian. First of all, in order to deal reliably with the effects of disorder resulting from the random substitution of $In^{3+}$ by $Mn^{2+}$ and treat non perturbatively the coupling between localized $Mn^{2+}$ spins ($S=5/2$) and itinerant holes, we perform an exact real space diagonalization (no effective medium is used) of the following Hamiltonian model\cite{richard-epl2007,richard1}: 
\begin{eqnarray}
H_{\rm V-J}=-\sum_{i,j,\sigma} t_{ij} c^{\dagger}_{i\sigma}  c^{}_{j\sigma} + \sum_{i}  J_{i} {\bf S}_{i}\cdot{\bf s}_{i} + \sum_{i\sigma}  V_{i}c^{\dagger}_{i\sigma}  c^{}_{i\sigma}
\label{Hamiltonian-VJ}
\end{eqnarray}
In this way crucial features such as localization effects of the itinerant carriers  are properly included in the Mn-Mn couplings.
\begin{figure}[t]\centerline
{\includegraphics[width=3.40in,angle=0]{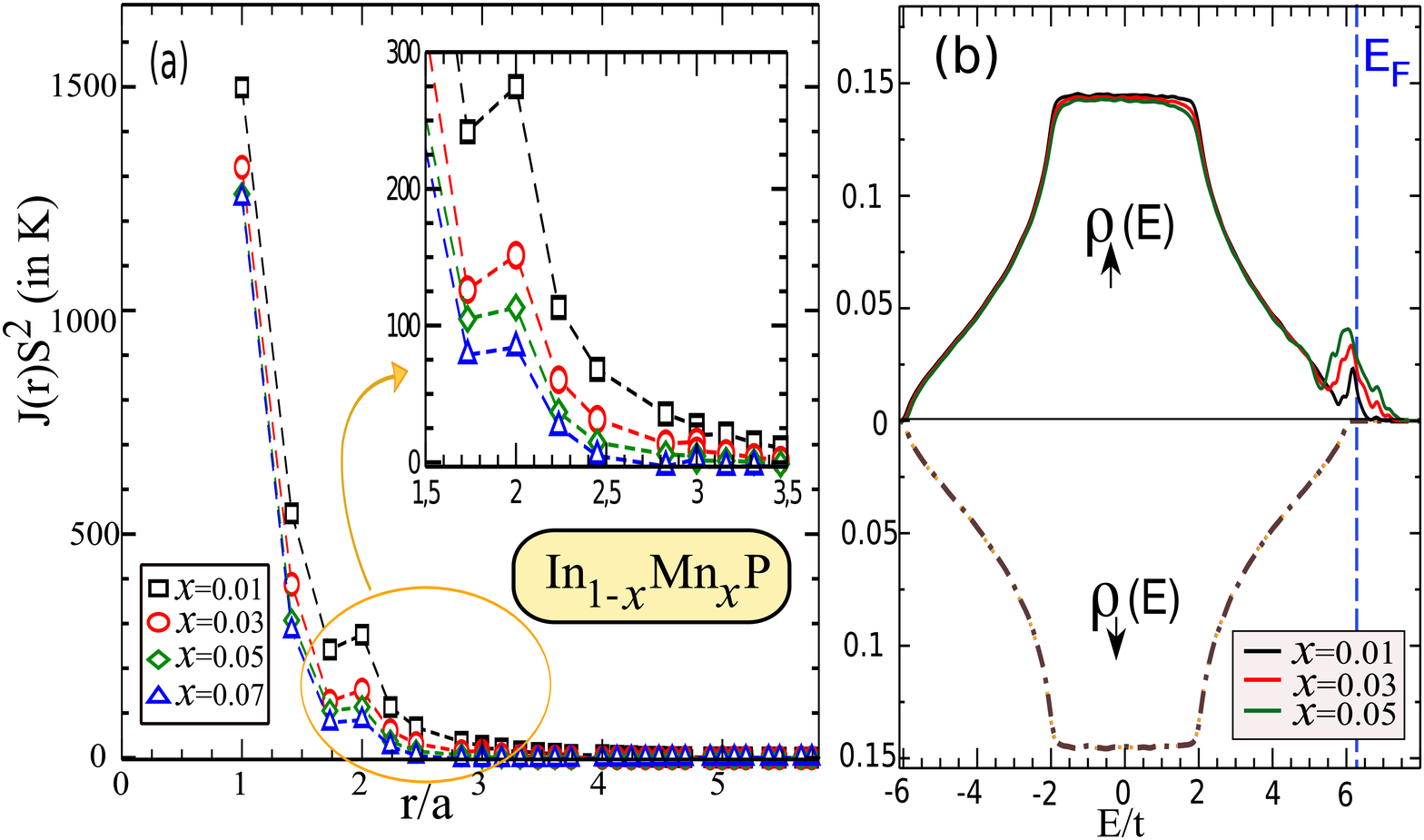}}
\caption{(Color online) (a) Couplings $J(r)S^2$ (in $K$) in In$_{1-x}$Mn$_{x}$P as a function of Mn-Mn distances for various $x$. Calculations are performed for well annealed samples (1 hole/Mn) using KPM on a system of size $72^3$. Inset: expanded view. (b) Spin resolved density of states, dashed vertical line indicates the Fermi level.}
\label{fig1}
\end{figure} 
In the first term ($i$,$j$) runs over all lattice sites, in the second and third ones the sum is restricted to Mn occupied sites only.
This model depends on 3 physical parameters ($t$,$V$,$J$) discussed thereafter.
For the sake of simplicity, we restrict to nearest neighbour hopping only, thus $t_{ij}$=$t$  when $i$ and $j$ are nearest neighbour.
$c^{\dagger}_{i\sigma}$  ($c^{}_{i\sigma}$) is the creation (annihilation) operator of a hole with spin $\sigma$ at site $i$. 
 $J_{i}=J$  is the $p$-$d$ coupling between the localized Mn spin ${\bf S}_{i}$ at site $i$ and the itinerant hole, its quantum spin operator is ${\bf s}_{i}$.
The on-site spin independent scattering potential $V_{i}=V$ results from the substitution of a host cation by a transition metal ion at site $i$.
It should be stressed that this V-J Hamiltonian can be derived from the more general Anderson Hamiltonian. The canonical Schrieffer-Wolff\cite{Schrieffer1966} transformation applied to the Anderson Hamiltonian reveals two particular terms: ($i$) the $s,p$-$d$ part that describes the magnetic interaction between carriers and localized spins and ($ii$) a non-magnetic scattering term. This latter term, often neglected in the literature, corresponds to the on site potential $V$. In order to reproduce the density of state heavy hole effective mass of the host semiconductor, the hopping term has been set to $t=0.7\,\mathrm{eV}$. We expect the amplitude of $J$ to be close to that of Mn doped GaAs, thus it is set to $1.2\,\mathrm{eV}$\cite{Okabayashi}. The last remaining parameter $V$ is chosen in order to reproduce the specific position of the hybridized $p$-$d$ acceptor level\cite{richard1} in InP ($220\,\mathrm{meV}$). This leads to $V=V_{\rm InP}$=$2.4t$ close to that of Mn doped GaAs, $V_{\rm GaAs}=1.8t$. Note that this model has already been successful to give an overall understanding of both magnetic and transport properties of the prototypical (Ga,Mn)As alloy\cite{richard1}. Yet at this stage, there is no guarantee that this model also applies to (In,Mn)P.
For a given Mn concentration and each spatial disorder configuration, the V-J Hamiltonian Eq.(\ref{Hamiltonian-VJ}) is used in combination with the Kernel Polynomial method (KPM)\cite{KPM,Lee}.
Within this recursive method the Heisenberg Hamiltonian is approximated by a truncated series of Chebyshev polynomials of order $n$. 
The computed expansion coefficients (moments of order $n$) allow for a fast calculation of the magnetic exchange integrals $J_{ij}$. In order to minimize finite size effects and statistical errors, the calculations of the couplings are made on a sufficiently large system ($72^3$ sites) and an average over few hundred disorder configurations is also performed.
\begin{figure}[t]\centerline
{\includegraphics[width=3.4in,angle=0]{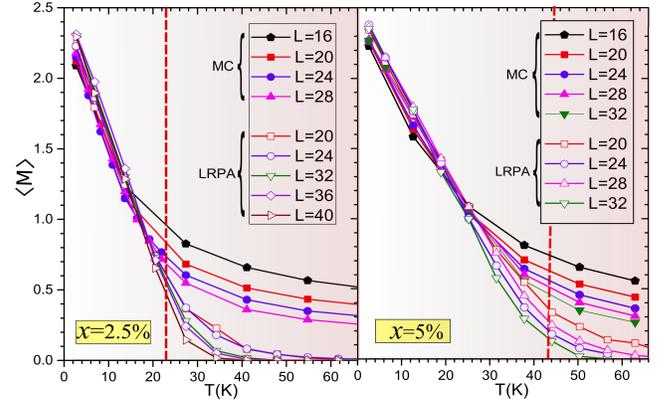}}
\caption{(Color online) Averaged magnetizations as a function temperature for two Mn concentrations, $x=0.025$ and $x=0.05$, and various system sizes. The calculations are performed within both LRPA and MC simulations. The vertical red dashed lines indicate the measured Curie temperature.}
\label{fig2}
\end{figure} 
In Fig.~\ref{fig1} both spin resolved density of states (DOS) and Mn-Mn couplings $J(r)S^2$ are depicted for impurity concentration $x$ ranging from $0.01$ to $0.07$. In the DOS we clearly see that in (In,Mn)P itinerant carriers are always fully polarized, and the Fermi level lies in a well defined impurity band (not totally separated from the valence band).
\begin{figure*}[t]\centerline
{\includegraphics[width=6.10in,angle=0]{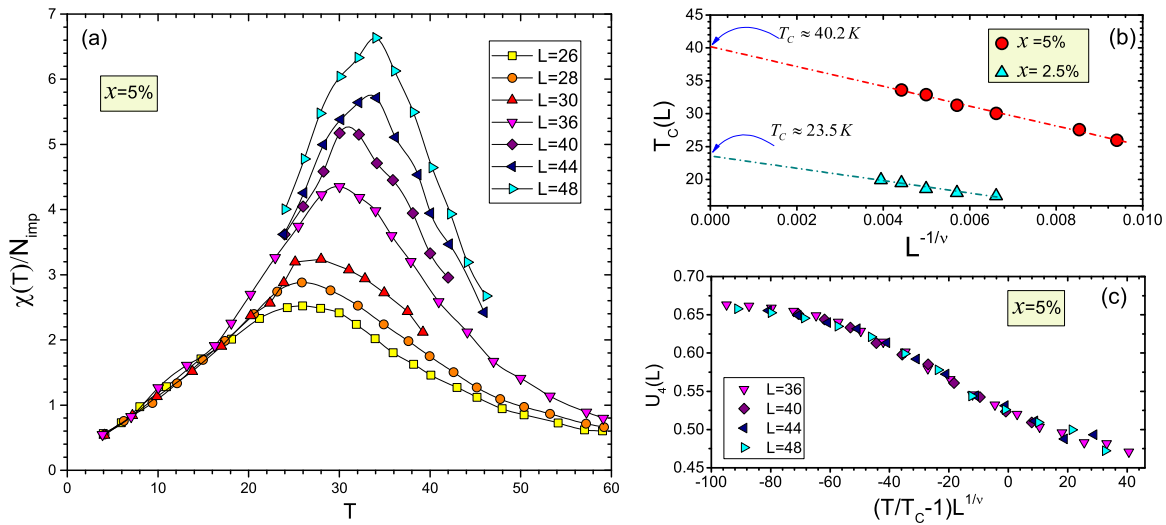}}
\caption{(Color online)
(a) Monte Carlo calculations of the averaged spin susceptibility $\chi(T)$ per impurity as a function of temperature for various system sizes $N=L^{3}$, where $L=26,...,48$.  The Mn concentration is $5\%$. (b) Finite size scaling of $T_C(L)$ (extracted from the maximum of $\chi(T)$) as a function of $L^{1/\nu}$ with $\nu=0.714$ for $x = 2.5$ and $5\%$. (c) Collapse plot for the Binder cumulant $U_4(L)$ as a function of $L^{1/\nu}(\frac{T}{T_{C}}-1)$ with $T_C=40.2\,\mathrm{K}$ and for $x=5\%$.}
\label{fig3}
\end{figure*} 
Some preliminary results for both the Drude weight (DC conductivity) and the typical density of states indicate that at the Fermi level carriers are localized, thus the compounds are insulating for the whole Mn concentration range\cite{rb-gb-preparation}. This agrees well with the insulating behaviour reported experimentally\cite{Zhou-R2014,Khalid-2015}. However, this is in contrast with the case of Mn doped GaAs that can exhibit an insulator to metal transition after annealing.
The exchange integrals shown in Fig.~\ref{fig1}\textcolor{blue}{(a)} are clearly not of standard oscillating RKKY type, since they are essentially ferromagnetic at all distances and rather short ranged. This feature is similar to what can be seen in (Ga,Mn)As\cite{EPJB2011} and results from the position of the Fermi level lying in the impurity band. However the couplings are shorter range in (In,Mn)P. As the Mn concentration increases the couplings at relevant distances are strongly reduced due to the increase of the multiple hole-impurity scatterings. Thus, so far, it is not clear whether the critical temperature will increase or not with the impurity concentration.

We now describe the second step of our approach, namely the determination of the magnetic properties. We consider the three dimensional dilute Heisenberg model: 
\begin{eqnarray}
H_{\rm Heis}=-\sum_{i,j} J_{ij} {\bf S}_{i}\cdot{\bf S}_{j}
\label{Hamiltonian-Heisenberg}
\end{eqnarray}
It describes the effective magnetic interactions between the localized Mn spins (${\bf S}_{i}$ are 3 dimensional normalized spin vectors and the sum runs over the impurity sites only).
The Mn concentration and hole density dependent exchange integrals are those shown in Fig.~\ref{fig1}.
To insure a proper treatment of the percolation effects, the magnetic excitations localization and the thermal/transverse fluctuations, we use classical Monte Carlo (MC) simulations and/or local Random Phase approximation (LRPA)\cite{EPL2005} to compute the magnetic properties of the system.

Let us now discuss the results of these calculations. In Fig.~\ref{fig2} we show the magnetization curves as a function of temperature for two Mn concentrations ($x=2.5$ and $5 \%$). The calculations have been performed for various system sizes, ranging from $16^3$ to $40^3$.
At low temperatures, below $15\,\mathrm{K}$ for $2.5\%$ doped and $30\,\mathrm{K}$ for $5\%$ doped compounds, the magnetization curves obtained within both methods are very close to each other. It reveals the same kind of highly unconventional linear decrease with temperature. This is in clear contrast with the conventional Brillouin like shape for which the magnetization has a relatively weak $T$-dependence at low temperature. Our findings are in qualitative agreement with measured data (SQUID and XMCD of Ref.\cite{Zhou-R2014}). In what follows, we will provide a quantitative comparison. As seen for the largest size, within  LRPA,  the linear decrease persists almost up to $T_C$. The slope (in absolute value) roughly decreases by a factor of two when the Mn concentration is doubled. This unconventional shape of the magnetization curve must be ascribed to the two concurring effects of disorder/dilution and very short range magnetic couplings. (Ga,Mn)As is at the edge that separates the conventional and unconventional behaviours.
In (In,Mn)P, it is worth noting that the weak extent of these couplings is a consequence of the localization of the carrier wave functions in this diluted insulating compound.
For high temperatures, we observe a much slower decay of the magnetization in the Monte Carlo simulations compared to LRPA results. The MC averaged magnetizations are more sensitive to finite size effects. An accurate determination of the critical temperature using the MC magnetization data is therefore difficult, especially in the dilute regime. On the other hand, LRPA allows for a direct calculation of the Curie temperature using a semi analytical formula\cite{EPL2005}, and leads to $T_{C}= 23\,\mathrm{K}$ and $43\,\mathrm{K}$ for $x= 2.5$ and $5\%$ respectively. Proceeding this way finite size effects are in practice very weak\cite{NJP2010,PRL2008} and the convergence of the calculations is very fast. 
On the other side, Monte Carlo simulations need far more efforts and CPU time to find accurate values of $T_C$. For instance, an often used method is based on the computation of the fourth order Binder cumulants\cite{MC-ref} $U_4(L)=1-\dfrac{\langle m^{4}\rangle}{3\langle m^{2}\rangle^{2}}$, where $m$ is the magnetization of a given system of size $L$. For various and sufficiently large sizes, the crossing point should occur at $T=T_C$. In our case, because of finite size effects, the crossing is not so accurate but occurs between $37\,\mathrm{K}$ and $43\,\mathrm{K}$ for the concentration $x=5\%$. A more precise result would require more computing resource, especially in the dilute regime when the couplings are short ranged. Alternatively we use some renormalisation group scaling arguments in order to extract $T_C$ in the thermodynamic limit. For the three dimensional Heisenberg model the Harris criterion\cite{Harris-1974} predicts that the critical exponents remain unaltered by disorder. Violation of this criterion has often been quoted in the literature. However, for finite system sizes, non universal corrections to the finite size scaling are to be considered in order to suppress this apparent violation as demonstrated within large scale Monte Carlo simulations\cite{Cruz-1986,Beach-2005,Priour-2010}. 
Consequently we still have to use the critical exponents of the pure 3D Heisenberg model. In order to extract $T_C$, we study the behaviour of $T_C(L)$, that corresponds to the temperature of the maximum of the spin susceptibility $\chi(L)$. In the thermodynamic limit the correlation length $\xi$ is comparable to the system size $L$, with the assumption  $\xi(T_C(L)-T_C) = aL$ (where $a$ is a constant). Since the asymptotic scaling is $\xi \propto(T-T_C)^{-\nu}$, it is thus expected that $T_C(L) - T_C \propto L^{-1/\nu}$, where $\nu = 0.714$ is the correlation length critical exponent\cite{MC-ref}. The real $T_C$ is then extrapolated from the finite size scaling analysis of $T_C(L)$. 
In Fig.~\ref{fig3}\textcolor{blue}{(a)} we have plotted the Monte Carlo spin susceptibility $\chi(T)=xL^3(\langle m^{2}\rangle - \langle m\rangle^{2})$ as a function of the temperature for $x = 5\%$ and for sizes ranging from $L=26$ to $L=48$. The finite size scaling of $T_C(L)$ as a function of $L^{-1/\nu}$ is depicted in Fig.~\ref{fig3}\textcolor{blue}{(b)} where the best linear fits lead to $T_C= 40.2\,\mathrm{K}$ and $23.5\,\mathrm{K}$ for $x=5$ and $2.5\%$ respectively.
\begin{figure}[t]\centerline
{\includegraphics[width=3.40in,angle=0]{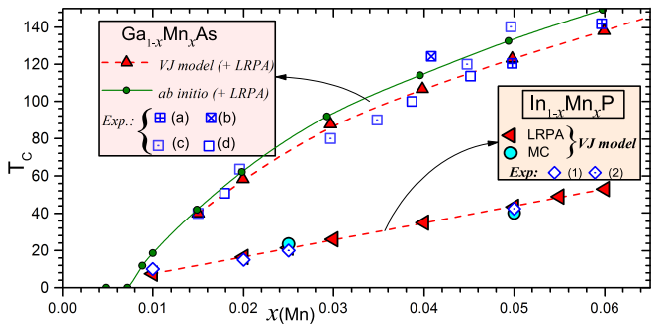}}
\caption{(Color online) Measured and calculated (MC and LRPA) ${\rm T}_{\rm C}$ (in K) as a function of Mn concentration $x$ in (In,Mn)P and (Ga,Mn)As. The green continuous curve is obtained from first principles combined with LRPA (from Ref.\onlinecite{compens}), red up triangles from the VJ model (Ref.\onlinecite{EPJB2011}). (In,Mn)P (this work): MC (circles) and LRPA (left triangles). Experimental data (2) and (1) are from Ref.\onlinecite{Zhou-R2014} and \onlinecite{Khalid-2015} respectively, and Refs.  \onlinecite{edmonds1,chiba,jungwirth,stone} for (a), (b), (c) and (d) respectively.
}
\label{fig4}
\end{figure}
These values of $T_C$ were used to check the compatibility with the scaling collapse plot for the Binder cumulants (as seen in Fig.~\ref{fig3}\textcolor{blue}{(c)} for $x=5\%$). Moreover both values agree quite well with those obtained within the LRPA approach. The variation of the Curie temperature as a function of the Mn concentration $x$ (from $1$ to $6 \%$) is shown in Fig.~\ref{fig4}.  Except from the two Mn concentrations 2.5 and 5\%, the other values of $T_C$ are calculated within the LRPA approach only. Note that these critical temperatures vary almost linearly with $x$ and the agreement with the recent available experiment data\cite{Zhou-R2014,Khalid-2015} is excellent for the whole range of Mn concentration. This success pleads in favour of our theoretical model which takes into account both the carrier mediated ferromagnetism and the correct location of the Mn binding energy in the host semiconductor. It should be mentioned that the standard virtual crystal approximation  would have led to largely overestimated critical temperatures (typically 1 order of magnitude larger at least!). In addition, the calculated $T_C$ in well annealed samples of Mn doped GaAs is also shown for comparison. 
For this compound we have reproduced previous results computed within both the VJ-model (with $t=0.7\,\mathrm{eV}$, $JS=4.3t$, $V=1.8t$) and ab initio based study\cite{compens} combined with LRPA. Experimental data from different groups\cite{edmonds1,chiba,jungwirth,stone} are also plotted. For (Ga,Mn)As, the V-J model also leads to very good agreement with both ab initio based studies and experiments. It is interesting to observe that the linear dependence found in (In,Mn)P is in strong contrast with the square root behaviour found in (Ga,Mn)As for which $T_C= A(x-x_c)^{1/2}$ (continuous line in Fig.~\ref{fig4}).

\begin{figure}[t]\centerline
{\includegraphics[width=3.40 in,angle=0]{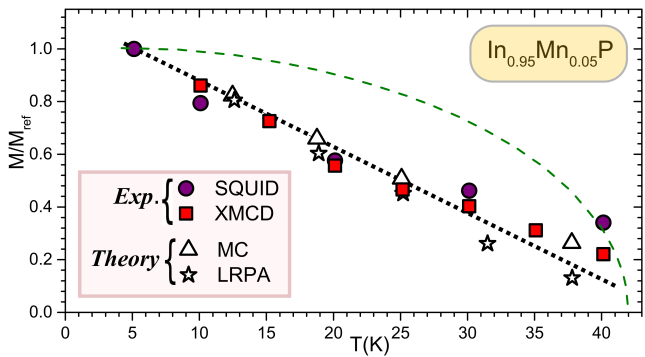}}
\caption{(Color online) Measured (SQUID and XMCD from Ref.\onlinecite{Zhou-R2014}) and calculated (MC and LRPA) magnetization as a function of the temperature in Kelvin for the Mn dilution $x=5\%$. $M_{\rm ref}$ is a magnetization reference at $T=5\,\mathrm{K}$. The dashed curve is the standard Brillouin magnetization for $S=5/2$.
}
\label{fig5}
\end{figure}

Finally, we propose to compare directly our calculated magnetization curves (MC and LRPA for the largest size $L=32$) with the measured ones (SQUID and XMCD) for the Mn concentration $x=5\%$. The experimental data are  extracted from Ref.[\onlinecite{Zhou-R2014}]. 
To facilitate discussion the magnetizations are divided by a reference magnetization $M_{\rm ref} = M(T=5\,\mathrm{K})$. The relative magnetizations as a function of $T$ are shown in Fig.~\ref{fig5}. This figure underlines the highly unconventional non-Brillouin-function like character of the magnetization curve (as a reminder the $S=5/2$ Brillouin magnetization is also shown in Fig.~\ref{fig5}).
An overall good quantitative agreement can be observed from low temperatures up to roughly $70\%$ of the experimental critical temperature $T_C\sim 42\,\mathrm{K}$. The almost linear slope of this unconventional magnetization curve is well reproduced. One should note that such a behaviour has never been observed in the case of well annealed Mn doped GaAs, the magnetization is closer to the standard Brillouin shape\cite{Chen,Deng}.

To conclude, starting from the V-J model and relying on efficient calculation techniques, such as Kernel Polynomial Method, Monte Carlo simulations and Local Random Phase Approximation we have addressed the nature of the ferromagnetism in Mn doped InP. Although the host InP is very similar to GaAs (gap, effective masses) their magnetic properties appear to be drastically different. The critical temperature in (In,Mn)P is much smaller than that of (Ga,Mn)As. Furthermore it scales linearly with Mn concentration in (In,Mn)P, in contrast to the square root behaviour found in (Ga,Mn)As. Moreover the magnetization curve exhibits an unconventional shape in (In,Mn)P, it varies almost linearly with temperature, in contrast to the conventional shape of well annealed (Ga,Mn)As samples. Our study reveals that the origin of these drastic changes of the magnetic properties is the extreme sensitivity to the  position of the Mn acceptor level. Our calculations are in excellent agreement with recent measurements for both critical temperatures and temperature dependent magnetization. Our results are only consistent with a Fermi level lying in an impurity band, it rules out the possibility to understand the physical properties of (In,Mn)P within the valence band scenario. More detailed experimental results for other III-V DMS compounds would be of great interest in order to improve our understanding of the magnetic properties in these systems.

\begin{acknowledgments}
D.M. and U.L. thank the Martin-Schmeißer-Stiftung for financial support. S. Z. acknowledges the support by Helmholtz-Association (VH-NG-713).
\end{acknowledgments}

\end{document}